\documentclass[journal,twoside,web]{ieeecolor}
\usepackage{tmi}
\usepackage{cite}
\usepackage{amsmath,amssymb,amsfonts}
\usepackage{algorithmic}
\usepackage{graphicx}
\usepackage{textcomp}
\usepackage{multirow}
\usepackage{cleveref}
\usepackage{caption}
\usepackage[T1]{fontenc}

\usepackage[utf8]{inputenc}
\usepackage{newunicodechar}
\usepackage[normalem]{ulem}
\usepackage{array,booktabs,tabularx}
\usepackage{booktabs,multirow,graphicx}
\usepackage{etoolbox}

\usepackage{arydshln}
\newcommand{\dotrule}{\noalign{\vskip 1.5pt}\hdashline\noalign{\vskip 1.5pt}}

\newunicodechar{−}{\textminus}

\def\BibTeX{{\rm B\kern-.05em{\sc i\kern-.025em b}\kern-.08em
    T\kern-.1667em\lower.7ex\hbox{E}\kern-.125emX}}
\begin{document}
\title{DiffUltra: Task-Guided Pathology Synthesis for Large-Scale Real-World Lung Ultrasound Data}
\author{Gary Y. Li, Yu-Cheng Chou, Li Chen, Mohsen Zahiri, Priyam Patel, Balasundar Raju, Jochen Kruecker, and Alvin Chen
\thanks{All authors are or were with Philips North America, Cambridge, MA 02141 USA (e-mail: \{ye.li, li.chen\_1, priyam.patel, balasundar.raju, jochen.krucker, alvin.chen\}@philips.com). Y.-C. Chou is now with the Department of Computer Science, Johns Hopkins University, Baltimore, MD 21218 USA (e-mail: ychou11@jhu.edu). M. Zahiri is now with Intuitive Surgical, Sunnyvale, CA 94086 USA (e-mail: mohsen.zahiri2@gmail.com).}
\thanks{The study was approved by Western-Copernicus Group (WCG) as the central Institutional Review Board (IRB; protocol 20223553), with local approval at each participating site. Written informed consent was obtained from all participants or their legally authorized representatives.}
\thanks{This study was funded in part by the U.S. Department of Health and Human Services; Administration for Strategic Preparedness and Response; Biomedical Advanced Research and Development Authority (BARDA), under contract 75A50120C00097. The contract and federal funding are not an endorsement of study results, product or company.}
\thanks{AI was used to assist with manuscript revision and literature review. All AI-assisted content was reviewed, verified and approved by the authors.}}
\maketitle

\begin{abstract}
Despite their scale, large medical imaging datasets may underrepresent clinically relevant pathological presentations. Synthetic data can expand coverage, but findings must remain compatible with surrounding anatomy. This is challenging in ultrasound, where findings resemble surrounding tissue and must be positioned relative to anatomical landmarks. We propose DiffUltra, a task-guided latent diffusion pipeline that models pathological structure and spatial context, and screens synthetic images for downstream utility. DiffUltra conditions generation on structural masks decomposing findings into anechoic, hypoechoic, and hyperechoic subregions. Two placement strategies insert findings into pathology-absent backgrounds or replace existing findings with structurally matched pathology from other patients. A filter trained exclusively on real data retains images whose predictions agree spatially with target boxes. We evaluate DiffUltra for lung ultrasound detection of pleural effusion and consolidation, using over 700,000 expert-annotated frames from 598 patients at 12 sites. Averaged across three real-data scales, synthetic insertion significantly improved video-level AUROC for both pleural effusion (\textit{p}\,=\,0.017) and consolidation (\textit{p}\,$<$\,0.01). For pleural effusion, frame-level average precision improved by 4.5 percentage points on average and by 6.1 points using only 10\% of real data (\textit{p}\,$<$\,0.01 for both). Synthetic insertion also yielded larger average precision gains than an existing diffusion-based lesion-insertion method at every scale. At full scale, targeted synthesis improved AUROC for both pathologies by 1.1--3.9 points in an underrepresented anatomical zone and on a held-out acquisition system. These results suggest that anatomically constrained, task-guided synthesis can help address coverage gaps in ultrasound training data.
\end{abstract}


\section{Introduction}
\label{sec:introduction}

\IEEEPARstart{R}{eliable} medical imaging AI requires training data that reflects the clinically relevant diversity of pathological presentations. Even large, annotated datasets can incompletely capture anatomical and acquisition-related variation. Consequently, models may perform well in aggregate yet fail on cases underrepresented or absent in the training data \cite{oakdenrayner2020hidden}.

Standard class-imbalance methods, such as oversampling, cost-sensitive learning, and ensembling, reweight or resample existing examples \cite{salmi2024handling}, \cite{chawla2002smote} but cannot recover variation absent from the training data. Generative image synthesis offers a mechanism to address these coverage gaps: generative adversarial networks (GANs) and diffusion models can produce visually realistic medical images \cite{goodfellow2014gan}, \cite{ho2020denoising} and improve performance in data-limited settings \cite{ktena2024generative}. Yet visual realism, as assessed through perceptual similarity metrics, does not guarantee downstream utility \cite{xing2023you}. A visually convincing foreground finding can be anatomically or spatially incompatible with the background, potentially degrading downstream performance. Effective synthetic augmentation therefore requires targeting underrepresented cases with findings that are compatible with surrounding anatomy and verified against the downstream task.

Foreground--background compatibility is particularly challenging in ultrasound. Unlike CT tumors, which are often texturally distinct from healthy tissue \cite{chen2024towards}, ultrasound pathologies often present as subtle echogenicity variations that blend with the surrounding anatomy \cite{yang2022current}. Furthermore, because the field of view depends on probe position and orientation \cite{yang2022current}, findings cannot be localized reliably using fixed image coordinates and instead must be positioned relative to anatomical landmarks. Thus, in ultrasound, a synthesized finding must preserve its internal structural characteristics while remaining spatially consistent with the anatomy visible in the imaging view.

% ===================== Figure I =====================
\begin{figure*}[!t]
\centering
\includegraphics[width=\textwidth]{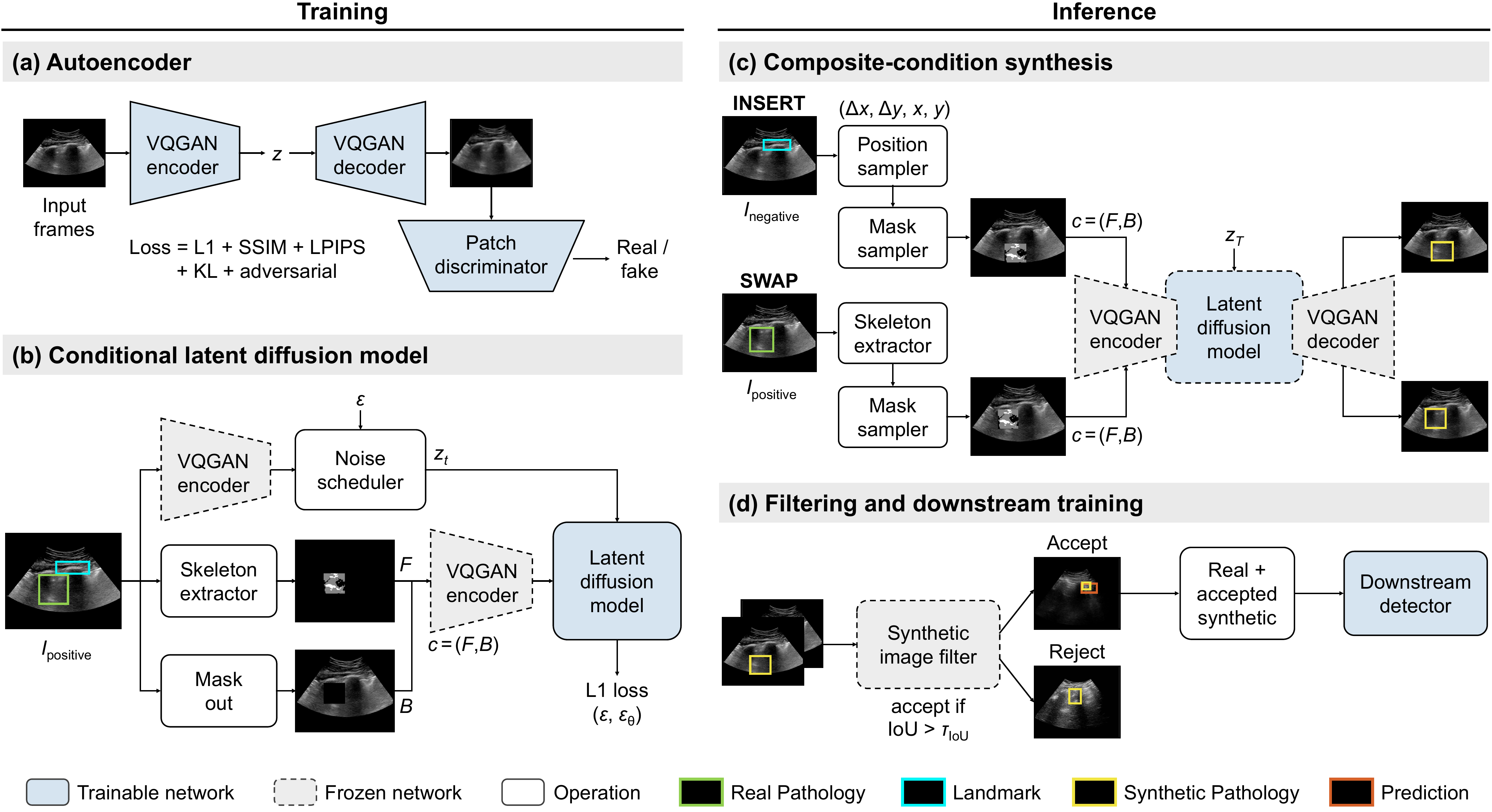}
\caption{DiffUltra pipeline. (a) VQGAN autoencoder training. (b) Latent diffusion model training on foreground \(F\) and background \(B\) conditions from real annotated images. (c) INSERT or SWAP combines a retrieved skeleton mask with a host image to form the composite condition \(c = (F, B)\), from which a frozen diffusion model and VQGAN decoder synthesize an image. (d) The synthetic image filter screens synthesized images, and accepted images are used for detector training.}
\label{fig:diffUltra_trainPipeline}
\end{figure*}

We present DiffUltra, a synthesis framework for ultrasound that addresses foreground--background compatibility across three stages: structural representation, contextual positioning, and task-based filtering (Fig.~\ref{fig:diffUltra_trainPipeline}). Structural representation captures the internal echogenic organization of the finding, contextual positioning controls the finding's spatial relationship to anatomical landmarks, and task-based filtering provides an empirical assessment of the resulting compatibility using a detector trained on real data. This synthesis capability can then be directed toward coverage gaps in the training data through targeted augmentation. The present study extends our prior work \cite{chou2025ultrasound}, which first proposed a diffusion-based framework for ultrasound synthesis with positional awareness. Here, we extend the framework in three ways: refining the structural representation to capture anechoic, hypoechoic, and hyperechoic subregions; incorporating complementary insertion and replacement strategies to address distinct coverage gaps; and introducing a task-based synthetic image filter (SIF) that retains images passing a detection-based criterion learned from real data. To our knowledge, this combination of structural representation, contextual positioning, and task-based filtering has not previously been applied to pathology synthesis in ultrasound or any other modality.

We apply the framework to lung ultrasound (LUS) pathology detection \cite{roy2020deep,tsai2021automatic,durrani2022automatic,li2025spatiotemporal}, characterizing the utility of synthetic augmentation as the scale of real training data increases. We evaluate two LUS pathologies, pleural effusion (PE) and consolidation (CON), across data scales spanning approximately 90,000 to 750,000 annotated frames (70 to 598 patients). Synthetic insertion improved detection of both pathologies on average across scales, with the largest gains when real training data were scarce. Improvements were larger and more consistent for PE than for CON. Targeting specific coverage gaps, such as a rare anatomical zone or an acquisition system absent from training, produced further gains at full data scale.

The main contributions of this work are as follows:

\begin{enumerate}
\def\labelenumi{\arabic{enumi}.}
\item
  A latent diffusion synthesis framework for ultrasound that explicitly addresses foreground--background compatibility through structural representation, contextual positioning, and task-based filtering.
\item
  A systematic characterization of synthetic augmentation across two ultrasound pathologies and multiple training data scales, with paired patient-level statistical comparisons against real-only training.
\item
  A targeted synthetic augmentation strategy for identified distributional coverage gaps, with additional gains for underrepresented subgroups at full data scale.
\end{enumerate}

\section{Related work}\label{related-work}

% Diffusion-based generative models have increasingly been used for medical image synthesis and augmentation \cite{goodfellow2014gan}, \cite{ho2020denoising}. For such applications, the challenge is not simply generating visually convincing images, but generating samples compatible with their context and useful for the downstream task.

\subsection{Foreground--Background Compatibility}\label{foregroundbackground-compatibility-in-pathology-synthesis}

A common strategy for controllable pathology synthesis decouples generation of a finding from its background. DiffTumor \cite{chen2024towards} models CT tumor shape as a deformed sphere and synthesizes texture within it. LeFusion \cite{zhang2025lefusion} combines a forward-diffused background with a reverse-diffused foreground for cardiac MRI. These approaches target modalities with relatively well-defined intensity and anatomical relationships, whereas ultrasound findings and surrounding anatomy often have similar echogenicity and indistinct boundaries \cite{yang2022current}.

Ultrasound synthesis methods have explored semantic, structural, and contextual conditioning. Liang et al. \cite{liang2022sketch} use fine-grained sketches for controllable generation, but customization requires labor-intensive mask editing. Stojanovski et al. \cite{stojanovski2023echo} use a coarsely sampled cone-shaped semantic mask, which is easier to generate but encodes less surrounding anatomy. Dahan et al. \cite{dahan2025csg} combine semantic and contextual conditioning for musculoskeletal ultrasound, using a reference image to guide texture without explicitly modeling a finding's position relative to anatomical landmarks. Anatomy-guided diffusion also improved fetal ultrasound classification and anomaly detection \cite{duan2025fetalflex}, but addresses whole-plane anatomy rather than placing a finding within an existing image.

For LUS, Heidari et al. \cite{heidari2026awdiff} use multi-scale wavelet and semantic conditioning to preserve fine diagnostic features such as B-lines and pleural-line irregularity, while Zhang et al. \cite{zhang2023diffusion} generate novel LUS variants from one or a few minority-class images. To our knowledge, none of these approaches explicitly models a finding's position relative to anatomical landmarks while enabling insertion into another patient's background. DiffUltra addresses this more specific challenge of foreground--background compatibility.

Our prior work \cite{chou2025ultrasound} modeled pathology position relative to anatomical landmarks using a conditional distribution learned from real patient data. Although the approach improved consolidation detection at one data scale, it used coarse conditioning, supported insertion only into pathology-absent images, and lacked automated assessment of downstream utility.

\subsection{Assessing Synthetic Image Utility}\label{assessing-synthetic-image-utility}

Realistic appearance does not guarantee downstream benefit. Xing et al. \cite{xing2023you} found that common image-quality metrics, including fidelity and diversity, correlate poorly with downstream performance. Similarly, Stojanovski et al. \cite{stojanovski2024efficient} found that visual realism did not necessarily predict segmentation or classification accuracy in echocardiography. In LUS, Fatima et al. \cite{fatima2025synthetic} reported that synthetic images were quantitatively similar to real images and largely indistinguishable to blinded experts, yet their inclusion did not improve downstream severity classification and sometimes reduced performance. By contrast, Zhang et al. \cite{zhang2023diffusion} improved LUS classification through minority-class diffusion synthesis, particularly when real minority-class data were scarce, despite weak correspondence between task benefit and conventional image-quality metrics. Related gains have been reported for diffusion-based ultrasound augmentation in fetal-plane classification \cite{tian2025enhancing}. Together, these findings show that synthetic-data utility depends on the method, context, and pathology, motivating direct task-based assessment rather than reliance on image realism alone.

Several methods filter or select synthetic data. Nazir et al. \cite{nazir2025diffusion} use a latent-space segmentation network to filter samples based on spatial intersection-over-union (IoU) with the intended localization. Gatoula et al. \cite{gatoula2026interpretable} assess utility at the dataset level using an interpretable utility-similarity measure, demonstrating utility-aware selection across imaging modalities, including ultrasound. In contrast, DiffUltra filters individual images based on spatial and confidence agreement between a detector trained on real data and the intended finding. This per-image task-based filtering is integrated with structurally and anatomically conditioned synthesis. DiffTumor \cite{chen2024towards}, Chou et al. \cite{chou2025ultrasound}, Dahan et al. \cite{dahan2025csg}, and Heidari et al. \cite{heidari2026awdiff} do not incorporate an equivalent task-based filtering step.

\subsection{Targeted Synthesis for Rare Subgroups}\label{targeted-synthesis-for-rare-subgroups}

Synthetic augmentation has long been used to address class imbalance. Frid-Adar et al. \cite{frid2018gan} used GAN-based synthesis for an underrepresented liver-lesion subtype, while copy-paste augmentation of rare object categories \cite{ghiasi2021simple} applies a related principle in computer vision. In LUS, Fatima et al. \cite{fatima2025synthetic} targeted underrepresented severity scores, but this did not improve downstream classification. These findings suggest that targeting imbalance alone may be insufficient without adequate anatomical and structural compatibility.

More recently, targeted synthesis has shown benefits at larger real-data scales. Chen et al. \cite{chen2025scaling} found that blanket synthetic augmentation plateaued as real-data volume increased, whereas targeting specific data gaps improved performance. Ktena et al. \cite{ktena2024generative} reported a similar pattern across histopathology, chest radiology, and dermatology: diffusion-based synthesis conditioned on demographic or acquisition attributes improved subgroup performance after general augmentation saturated. Their evaluation includes lung consolidation and pleural effusion, and their hospital-conditioned setting provides a relevant analogue to an out-of-distribution acquisition system. We investigate whether anatomically and structurally aware synthesis, rather than attribute-conditioned sampling, can address analogous coverage gaps in LUS.

% ===================== Figure II =====================
\begin{figure}[!t]
\centerline{\includegraphics[width=\columnwidth]{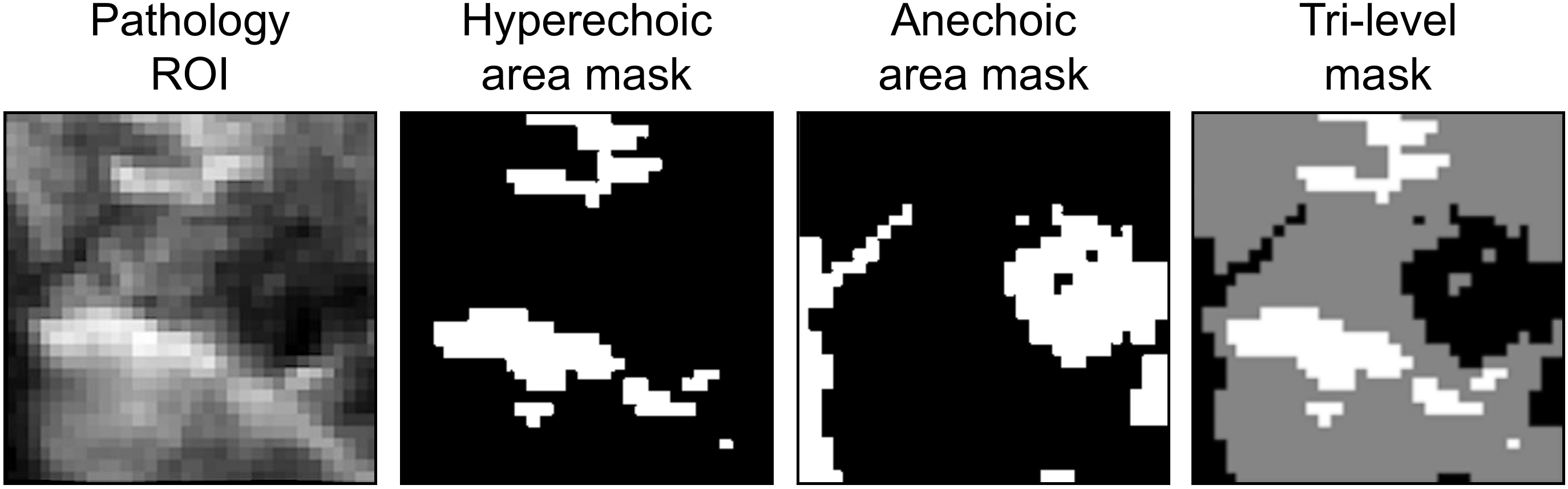}}
\caption{Tri-level structural mask construction. Left to right: pathology region, thresholded hyperechoic and anechoic areas, and the resulting mask used for conditioning and retrieval.}
\label{fig:skeleton_generation}
\end{figure}

% ===================== Figure III =====================
\begin{figure*}[!t]
\centering
\includegraphics[width=0.99\textwidth]{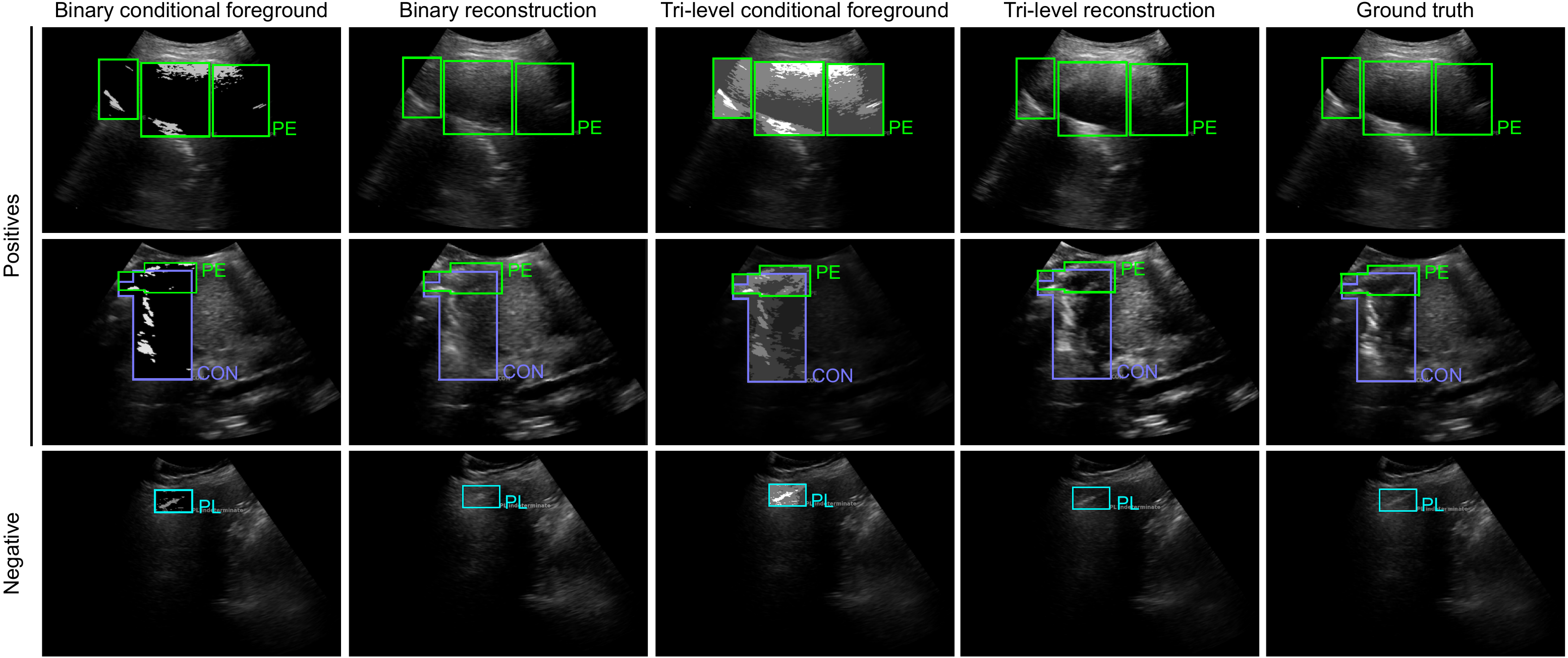}
\caption{Reconstructions from binary and tri-level conditional foreground masks, alongside ground truth images for pleural effusion (top), consolidation and pleural effusion (middle), and a negative image with only pleural line (bottom). Boxes: green, PE; purple, CON; cyan, pleural line (PL). Binary masks discard internal echogenic structure, which tri-level masks preserve.}
\label{fig:syn_img_vis}
\vspace{-4pt}
\end{figure*}

% ===================== Algorithm I =====================
\begin{figure}[!t]
\scriptsize
\setlength{\tabcolsep}{0pt}
\renewcommand{\arraystretch}{1.0}

\noindent
\textsc{Algorithm 1.} DiffUltra contextual positioning, sampling, and filtering.
\vspace{1pt}

\hrule
\vspace{2pt}

\noindent
\textit{Input:} Target image \(I\) (pathology-absent negative image \(I_{\text{negative}}\) or pathology-present positive image \(I_{\text{positive}}\)), skeleton dictionary \(\mathcal{D}\), thresholds \(\tau_{\mathrm{conf}}, \tau_{\mathrm{IoU}}\)

\vspace{1pt}
\noindent
\textit{Output:} Retained synthetic image \(I_{\text{syn}}\) or REJECT

\vspace{2pt}
\hrule
\vspace{2pt}

\begin{tabular}{@{}r@{\hspace{5pt}}p{\dimexpr\columnwidth-18pt\relax}@{}}

1:  & \textbf{if} Mode == INSERT \textbf{then} \\

2:  & \hspace{6pt}Locate landmark center \((x,y)\) in \(I_{\text{negative}}\) \\

3:  & \hspace{6pt}Sample offset \((\Delta x,\Delta y) \sim P(\Delta X,\Delta Y \mid X{=}x,Y{=}y)\) \textbf{(Eq.~\eqref{eq:position_offset})} \\

4:  & \hspace{6pt}Retrieve mask
      \(m \sim \mathrm{Uniform}(\mathcal{M})\)
      \textbf{(Eq.~\eqref{eq:mask_INSERT})} \\

5:  & \textbf{else if} Mode == SWAP \textbf{then} \\

6:  & \hspace{6pt}Extract target skeleton mask \(m_{\mathrm{orig}}\)
      from \(I_{\text{positive}}\) \\

7:  & \hspace{6pt}Construct candidate set
      \(\mathcal{M}(m_{\text{orig}})\) with size ratio \(\rho\)
      and threshold \(\tau_{\mathrm{Dice}}\)
      \textbf{(Eq.~\eqref{eq:mask_SWAP})} \\

8:  & \hspace{6pt}Retrieve mask
      \(m \sim \mathrm{Uniform}
      (\mathcal{M}(m_{\text{orig}}))\) \\

9:  & \textbf{end if} \\

10: & Construct composite condition \(c=(F,B)\)
      \textbf{(Eq.~\eqref{eq:composite_condition})} \\

11: & Initialize latent noise
      \(z_T \sim \mathcal{N}(0,\mathbf{I})\) \\

12: & \textbf{for} \(t=T\) down to \(1\) \textbf{do} \\

13: & \hspace{6pt}Compute
      \(z_{t-1} =
      \mathcal{S}
      (z_t,\varepsilon_\theta(z_t,t,c),t)\)
      \textbf{(Eq.~\eqref{eq:diffusion_denoise})} \\

14: & \textbf{end for} \\

15: & Decode final latent:
      \(I_{\text{syn}}=\mathrm{Dec}(z_0)\) \\

16: & Pass \(I_{\text{syn}}\) through SIF
      (detector trained on real data only) \\

17: & Extract predicted boxes with confidence
      \(> \tau_{\mathrm{conf}}\) \\

18: & Compute
      \(\overline{\mathrm{IoU}}_i(\tau_{\mathrm{conf}})\)
      for each ground-truth box \(G_i\)
      \textbf{(Eq.~\eqref{eq:SIF_IoU})} \\

19: & \textbf{if}
      \(\min_i \overline{\mathrm{IoU}}_i(\tau_{\mathrm{conf}})
      > \tau_{\mathrm{IoU}}\)
      \textbf{then} \\

20: & \hspace{6pt}\textbf{return}
      \(I_{\text{syn}}\) (Accepted) \\

21: & \textbf{else} \\

22: & \hspace{6pt}\textbf{return} REJECT \\

23: & \textbf{end} \\
\end{tabular}
\vspace{2pt}
\hrule
%\vspace{-10pt}
\end{figure}

% ===================== Figure IV =====================
\begin{figure*}[!t]
\centering
\includegraphics[width=\textwidth]{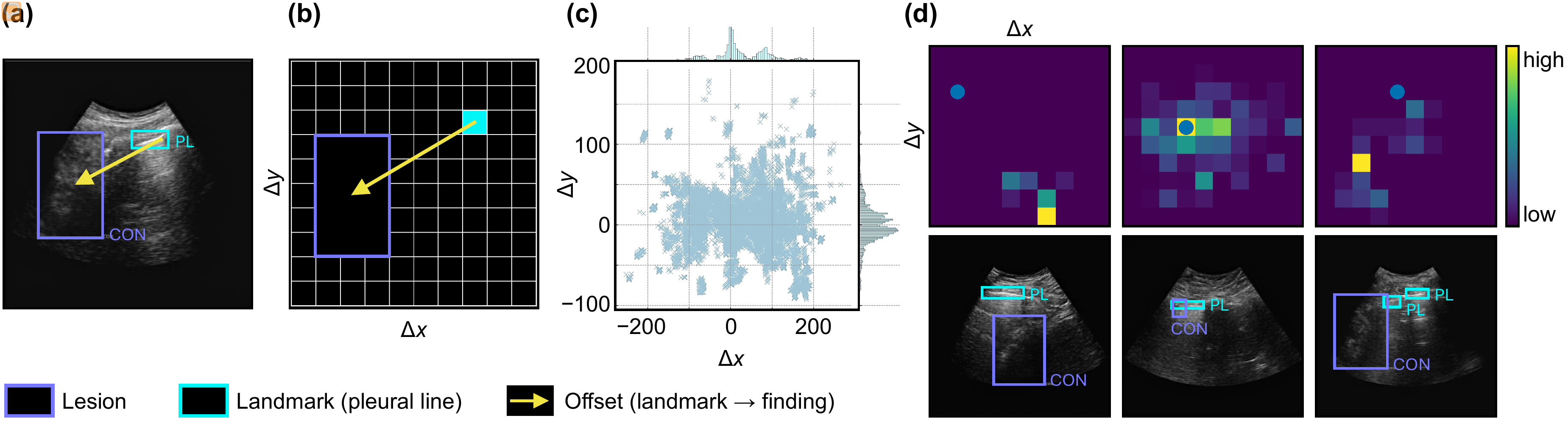}
\caption{Estimation of the conditional PMF used by the foreground position sampler. (a) Positional offset between a finding (CON) and its nearest anatomical landmark (PL). (b) Positional offset on the 10×10 sampling grid. (c) Offsets pooled across patients, probes, zones, and pathologies. (d) Conditional PMFs $P(\Delta X,\Delta Y \mid X,Y)$ for three landmark positions (blue dot).}
\label{fig:vis_PMFs}
\end{figure*}

\section{Methods}
\subsection{Overview}\label{overview}

DiffUltra addresses foreground--background compatibility through three sequential stages: structural representation of a finding, contextual positioning of the finding within an anatomical background, and task-based filtering of the resulting synthetic images. The pipeline separates generative model training from inference-time synthesis and filtering for downstream detector training (Fig.~\ref{fig:diffUltra_trainPipeline}; Algorithm 1).

\begin{itemize}
\item
  During training, an autoencoder with encoder \(\mathrm{Enc}(\cdot)\) and decoder \(\mathrm{Dec}(\cdot)\) learns a compressed latent representation \(z = \mathrm{Enc}(I)\). A latent diffusion model \(\varepsilon_{\theta}\) is then trained on frames annotated with ground-truth bounding boxes, conditioned on a composite condition \(c = (F,B)\) constructed from structural representations of real annotated findings (Section~\ref{structural-representation}).
\item
  During inference, each target image is paired with one of two contextual positioning strategies, INSERT or SWAP. These strategies construct \(c\) using anatomically plausible positions and structurally matched skeletons retrieved from real patient data (Section~\ref{contextual-positioning-insert-and-swap}). The diffusion model \(\varepsilon_{\theta}\) generates an image from \(c\), which the SIF (Section~\ref{task-based-synthetic-image-filter-sif}) accepts or rejects before the image is added to the real detector training set (Section~\ref{synthetic-augmentation-for-downstream-training}).
\end{itemize}

\subsection{Conditional Latent Diffusion Model}\label{conditional-latent-diffusion-model}

An autoencoder is first trained to map ultrasound frames into a lower-dimensional latent space, separating perceptual compression from diffusion modeling \cite{ho2020denoising}, \cite{rombach2022high}. Its encoder and decoder weights are frozen during diffusion model training.

The diffusion model operates in this latent space, conditioned on \(c = (F,B)\), with \(F\) and \(B\) encoded by the frozen encoder (Fig.~\ref{fig:diffUltra_trainPipeline}). Following \cite{ho2020denoising}, \cite{rombach2022high}, an image \(I\) with latent \(z_{0} = \mathrm{Enc}(I)\) is progressively noised over \(T\) steps:
\begin{equation}\label{eq:diffusion_noise}
\begin{gathered}
q(z_t \mid z_0) = \mathcal{N}\bigl(z_t;\, \sqrt{\bar{\alpha}_t}\, z_0,\, (1-\bar{\alpha}_t)\mathbf{I}\bigr),\\
\bar{\alpha}_t = \prod_{s=1}^{t}(1-\beta_s),
\end{gathered}
\end{equation}
and \(\varepsilon_{\theta}\) is trained to predict the injected noise given \(c\):
\begin{equation}\label{eq:diffusion_loss}
\mathcal{L}_{\text{LDM}} = \mathbb{E}_{z_0,\,\varepsilon\sim\mathcal{N}(0,\mathbf{I}),\,t}\bigl[\lVert \varepsilon - \varepsilon_\theta(z_t,t,c)\rVert_1\bigr].
\end{equation}
At inference, the latent is initialized as \(z_T \sim \mathcal{N}(0,\mathbf{I})\) and iteratively denoised:
\begin{equation}\label{eq:diffusion_denoise}
z_{t-1} = \mathcal{S}\bigl(z_t,\, \varepsilon_\theta(z_t,t,c),\, t\bigr), \qquad t = T,\ldots,1,
\end{equation}
where \(\mathcal{S}\) denotes the diffusion sampling step (e.g., denoising diffusion implicit models, DDIM \cite{song2021ddim}). Algorithm 1 summarizes inference. The final latent \(z_0\) is decoded to obtain \(I_{\text{syn}} = \mathrm{Dec}(z_0)\). Unlike unconditional or single-image-conditioned generation, separate foreground and background conditions allow the finding's structure and appearance to vary while retaining the anatomical context of the host image.

\subsection{Structural Representation}\label{structural-representation}

Conditioning on a finding's original texture would preserve its low-level appearance and limit diversity, whereas a binary mask would discard echogenic structure. DiffUltra instead uses a tri-level skeleton mask \(m\) that preserves geometric boundaries and coarse echogenicity while letting fine-grained texture vary. Given a pathology region of interest \(R\) cropped from an annotated bounding box, Otsu's method \cite{otsu1975threshold} progressively segments \(R\) into three echogenicity classes (Fig.~\ref{fig:skeleton_generation}):
\begin{equation}\label{eq:trilevel_mask}
m : R \rightarrow \{v_1, v_2, v_3\},
\end{equation}
where \(v_1\), \(v_2\), and \(v_3\) are assigned to anechoic, hypoechoic, and hyperechoic pixels, respectively. These classes capture acoustic structures relevant to the two LUS pathologies considered: anechoic regions represent fluid such as pleural effusion, hypoechoic regions capture consolidated tissue, and hyperechoic regions capture reflective structures such as the pleural line. This procedure is applied to every annotated finding to construct both the foreground condition \(F\) used in diffusion-model training and the skeleton dictionary \(\mathcal{D}\) used for retrieval. Explicitly representing these subregions is intended to reduce failure modes of binary conditioning, including hallucinated content within anechoic fluid regions and blurred boundaries within consolidated tissue. Fig.~\ref{fig:syn_img_vis} qualitatively compares reconstructions using binary \cite{chou2025ultrasound} and tri-level conditioning.

\subsection{Contextual Positioning: INSERT and SWAP}\label{contextual-positioning-insert-and-swap}

We introduce two strategies for combining a finding's structural representation with anatomical context. INSERT places a finding in a pathology-absent image, varying both its appearance and anatomical context. SWAP replaces an existing finding with a structurally matched finding from the dictionary, varying the finding while preserving the host context. Both strategies use a shared skeleton dictionary:
\begin{equation}\label{eq:skeleton_dictionary}
\begin{gathered}
\mathcal{D} = \{(m_k,\mathbf{p}_k)\}_{k=1}^{K},\\
\mathbf{p}_k =
\begin{cases}
(\Delta x_k,\Delta y_k,x_k,y_k), & \text{if available},\\
\varnothing, & \text{otherwise}.
\end{cases}
\end{gathered}
\end{equation}
The dictionary stores the skeleton mask \(m_k\) for each of \(K\) annotated findings and, when a key anatomical structure is present, its positional information \(\mathbf{p}_k\). INSERT requires positional information, whereas SWAP can use any entry.

\begin{enumerate}
\def\labelenumi{\alph{enumi})}
\item
  \textbf{INSERT:} Following \cite{chou2025ultrasound}, INSERT models the offset from the center of the nearest key anatomical structure \((x,y)\) to the finding center \((x',y')\) as \((\Delta x,\Delta y)=(x'-x,y'-y)\). The conditional probability mass function (PMF) \(P(\Delta X,\Delta Y\mid X,Y)\) is estimated empirically from real annotations by spatially binning the four coordinates \((\Delta x,\Delta y,x,y)\) using \(\psi(\cdot)\) (Fig.~\ref{fig:vis_PMFs}).
\end{enumerate}

The PMF and skeleton dictionary use the same binning function \(\psi(\cdot)\). At inference, given a landmark center \((x,y)\), an offset is sampled as:
\begin{equation}\label{eq:position_offset}
(\Delta x,\Delta y) \sim P(\Delta X,\Delta Y\mid X=x,Y=y).
\end{equation}
A skeleton mask \(m\) is then sampled uniformly from dictionary entries in the corresponding spatial bin:
\begin{equation}\label{eq:mask_INSERT}
\begin{aligned}
\mathcal{M} &= \bigl\{m_k \mid (m_k,\mathbf{p}_k)\in\mathcal{D},\ \mathbf{p}_k\neq\varnothing,\\
&\qquad \psi(\mathbf{p}_k)=\psi(\Delta x,\Delta y,x,y)\bigr\},\\
m &\sim \operatorname{Uniform}(\mathcal{M}).
\end{aligned}
\end{equation}
The retrieved mask is placed at its native size, centered at \((x+\Delta x,y+\Delta y)\), without rescaling.

\begin{enumerate}
\def\labelenumi{\alph{enumi})}
\setcounter{enumi}{1}
\item
  \textbf{SWAP:} Given an existing finding with mask \(m_{\text{orig}}\), SWAP samples a size- and structure-matched replacement uniformly from the candidate set. Let \(s_m=(w_m,h_m)\) be the size of mask \(m\), with \(s_{\text{orig}}\) the size of \(m_{\text{orig}}\):
\end{enumerate}
\begin{equation}\label{eq:mask_SWAP}
\begin{aligned}
\mathcal{M}(m_{\text{orig}}) &= \bigl\{m_k \mid (m_k,\mathbf{p}_k)\in\mathcal{D},\\
&\qquad \tfrac{1}{\rho} \leq s_{m_k} \oslash s_{\text{orig}} \leq \rho,\\
&\qquad \operatorname{Sim}(m_k,m_{\text{orig}}) \geq \tau_{\text{Dice}}\bigr\},\\
m &\sim \operatorname{Uniform}\bigl(\mathcal{M}(m_{\text{orig}})\bigr),
\end{aligned}
\end{equation}
where \(\oslash\) is elementwise division with componentwise inequalities, and \(\rho\geq1\) is a symmetric size tolerance. Similarity is the mean Dice score across \(L=3\) echogenicity levels:
\begin{equation}\label{eq:mask_similarity}
\operatorname{Sim}(m,m_{\text{orig}}) = \frac{1}{L}\sum_{l=1}^{L}\operatorname{Dice}\bigl(m^{(l)},m_{\text{orig}}^{(l)}\bigr),
\end{equation}
where \(m^{(l)}\) denotes the binary mask for echogenicity level \(l\).

\begin{enumerate}
\def\labelenumi{\alph{enumi})}
\setcounter{enumi}{2}
\item
  \textbf{Constructing the composite condition:} For each image, findings to be synthesized are processed independently via the INSERT or SWAP procedure, yielding masks \(m_{i}\) placed at target centers \((x_i', y_i')\) to form the foreground condition \(F\). For INSERT, each region is masked from the pathology-absent host image \(I_{\text{negative}}\) to form the background condition \(B\). For SWAP, we use \((x_i', y_i')\) = \((x_{\text{orig}}, y_{\text{orig}})\), and each region is masked from the pathology-present host image \(I_{\text{positive}}\).
\end{enumerate}

Let \(r_i \in \{0,1\}^{H \times W}\) denote the region to be removed at target location \((x_i',y_i')\), and let \(\widetilde{m}_i\) denote mask \(m_i\) embedded at that location and zero elsewhere. For \(N_g\) findings:
\begin{equation}\label{eq:composite_condition}
\begin{gathered}
F = \max_{i=1,\ldots,N_g} \widetilde{m}_i,\\
B = \Bigl(1 - \max_{i=1,\ldots,N_g} r_i\Bigr) \odot I_{\text{host}},
\end{gathered}
\end{equation}
where \(\odot\) is elementwise multiplication, and \(I_{\text{host}}\) is \(I_{\text{negative}}\) for INSERT and \(I_{\text{positive}}\) for SWAP. The composite condition \(c=(F,B)\) is passed to \(\varepsilon_{\theta}\), and the ground-truth box for each synthesized finding is \(G_i=\text{bounding-box}(m_i,x_i',y_i')\). During training, \(F\) and \(B\) are constructed from real annotated images.

\subsection{Task-Based Synthetic Image Filter (SIF)}\label{task-based-synthetic-image-filter-sif}

A detector trained exclusively on real data assesses spatial agreement between predicted boxes and the intended synthetic targets. The detector \(\varphi\) maps an image to predicted boxes and confidence scores, \(\varphi(I)=\{(b_j,s_j)\}\). The downstream detector uses the same architecture but is trained separately.

For each intended box \(G_i\), predictions \(\{(b_j,s_j)\}=\varphi(I_{\text{syn}})\) are filtered using the confidence threshold \(\tau_{\text{conf}}\), forming the set \(\mathcal{B}(\tau_{\text{conf}})=\{j\mid s_j>\tau_{\text{conf}}\}\). The average IoU with \(G_i\) is:
\begin{equation}\label{eq:SIF_IoU}
\overline{\text{IoU}}_i(\tau_{\text{conf}}) =
\begin{cases}
\tfrac{1}{|\mathcal{B}(\tau_{\text{conf}})|}\displaystyle\sum_{j\in\mathcal{B}(\tau_{\text{conf}})}\text{IoU}(G_i,b_j), & |\mathcal{B}(\tau_{\text{conf}})|{>}0,\\
0, & \text{otherwise}.
\end{cases}
\end{equation}
A synthetic image \(I_{\text{syn}}\) with \(N_{g}\) findings is accepted if:
\begin{equation}\label{eq:SIF_accept}
\text{Accept}(I_{\text{syn}}) = \mathbf{1}\Bigl[\min_{i=1,\ldots,N_g}\overline{\text{IoU}}_i(\tau_{\text{conf}}) > \tau_{\text{IoU}}\Bigr].
\end{equation}
Thus, every synthesized finding must satisfy the spatial-agreement criterion; a single failure rejects the entire image. A lower \(\tau_{\text{conf}}\) admits lower-confidence predictions, whereas a higher $\tau_\text{conf}$ considers only more confident predictions.

\subsection{Synthetic Augmentation for Downstream Training}\label{synthetic-augmentation-for-downstream-training}

Accepted synthetic images, for which \(\text{Accept}(I_{\text{syn}})=1\), are added to the real training set. Augmentation volume is controlled by the synthetic-to-real-positive ratio \(\gamma_{\text{syn}}=N_{\text{syn}}/N_{\text{real,+}}\), where \(N_{\text{real,+}}\) is the number of real pathology-positive training frames (Section~\ref{experimental-settings}). Augmentation is either untargeted, sampling uniformly across available background images, or targeted, applying INSERT and SWAP only to subsets corresponding to identified coverage gaps, such as underrepresented anatomical zones or acquisition systems.

% ===================== Table I =====================
\newcommand{\fnum}[1]{\hphantom{000,000}\llap{#1}}
\newcommand{\framecell}[2]{\begin{tabular}{@{}l@{\hspace{3pt}}r@{}}CON: & \fnum{#1}\\ PE: & \fnum{#2}\end{tabular}}
\newcommand{\framesingle}[1]{\begin{tabular}{@{}l@{\hspace{3pt}}r@{}}\phantom{CON:} & \fnum{#1}\end{tabular}}

\begin{table}[!t]
\centering
\footnotesize
\setlength{\tabcolsep}{4pt}
\renewcommand{\arraystretch}{1.15}
\caption{Data used for autoencoder pretraining on all available data and detector training at each real-data scale.}
\label{tab:2}
\begin{tabular*}{\columnwidth}{@{\extracolsep{\fill}}lrrrr@{}}
\hline\hline
\textbf{Stage} & \textbf{Subjects} & \textbf{Videos} & \textbf{PE frames} & \textbf{CON frames} \\
\midrule
Autoencoder pretraining & 2,110 & 59,176 & \multicolumn{2}{c}{5,229,141} \\
\dotrule
Detection (100\% scale) & 598 & 8,585 & 719,997 & 746,052 \\
Detection (50\% scale)  & 320 & 4,780 & 374,266 & 373,574 \\
Detection (10\% scale)  & 70  & 1,150 & 83,784 & 88,354 \\
\hline\hline
\end{tabular*}
\end{table}

% ===================== Table II =====================
\begin{table*}[!t]
\centering
\footnotesize
\caption{Frame-level detection (AP@0.5) and video-level classification (AUROC) with patient-level bootstrap 95\% CIs across 10\%, 50\%, and 100\% real-data scales, and averaged across scales. $\Delta$ is the change from the real-only baseline in percentage points; *$p < 0.05$ vs real-only, two-sided paired bootstrap. Bold marks the best value at each scale and averaged across scales.}
\label{tab:synthetic_utility_combined_ioup5}
\setlength{\tabcolsep}{3pt}
\renewcommand{\arraystretch}{1.08}
\setlength{\aboverulesep}{1pt}\setlength{\belowrulesep}{1pt}
\begin{tabular*}{\textwidth}{@{\extracolsep{\fill}}lcccccccc@{}}
\hline\hline
\multirow{2}{*}{\textbf{Experimental Setting}}
& \multicolumn{4}{c}{\textbf{PE}} & \multicolumn{4}{c}{\textbf{CON}} \\
\cmidrule(lr){2-5}\cmidrule(l){6-9}
& \textbf{AP@0.5} (95\% CI) & \textbf{$\Delta$}
& \textbf{AUROC} (95\% CI) & \textbf{$\Delta$}
& \textbf{AP@0.5} (95\% CI) & \textbf{$\Delta$}
& \textbf{AUROC} (95\% CI) & \textbf{$\Delta$} \\
\midrule
\multicolumn{9}{@{}l}{\textbf{\textit{10\% Real-data scale:}}} \\
Real only & 0.227 (0.177--0.270) & -- & 0.787 (0.755--0.818) & -- & 0.130 (0.097--0.167) & -- & 0.829 (0.806--0.853) & -- \\
DiffTumor \cite{chen2024towards} & 0.261 (0.206--0.306) & $+3.4$* & 0.806 (0.774--0.838) & $+2.0$* & 0.129 (0.099--0.165) & $-0.1$ & 0.834 (0.811--0.856) & $+0.5$ \\
DiffUltra SWAP & 0.220 (0.167--0.269) & $-0.6$ & 0.782 (0.750--0.812) & $-0.5$ & 0.111 (0.087--0.141) & $-1.9$ & 0.815 (0.789--0.840) & $-1.4$ \\
DiffUltra INSERT & \textbf{0.288 (0.225--0.342)} & $\mathbf{+6.1}$* & \textbf{0.819 (0.788--0.848)} & $\mathbf{+3.2}$* & \textbf{0.141 (0.106--0.183)} & $\mathbf{+1.0}$ & \textbf{0.851 (0.827--0.873)} & $\mathbf{+2.2}$* \\
\dotrule
\multicolumn{9}{@{}l}{\textbf{\textit{50\% Real-data scale:}}} \\
Real only & 0.372 (0.300--0.434) & -- & 0.870 (0.842--0.894) & -- & 0.258 (0.217--0.306) & -- & 0.884 (0.864--0.905) & -- \\
DiffTumor \cite{chen2024towards} & 0.385 (0.299--0.460) & $+1.3$ & \textbf{0.872 (0.848--0.896)} & $\mathbf{+0.3}$ & 0.253 (0.214--0.291) & $-0.5$ & 0.887 (0.867--0.905) & $+0.3$ \\
DiffUltra SWAP & 0.394 (0.299--0.474) & $+2.2$ & 0.857 (0.830--0.882) & $-1.3$ & 0.235 (0.196--0.269) & $-2.3$ & 0.881 (0.859--0.901) & $-0.3$ \\
DiffUltra INSERT & \textbf{0.399 (0.301--0.487)} & $\mathbf{+2.7}$ & 0.862 (0.832--0.890) & $-0.8$ & \textbf{0.270 (0.228--0.315)} & $\mathbf{+1.2}$ & \textbf{0.896 (0.876--0.915)} & $\mathbf{+1.2}$* \\
\dotrule
\multicolumn{9}{@{}l}{\textbf{\textit{100\% Real-data scale:}}} \\
Real only & 0.421 (0.358--0.473) & -- & 0.902 (0.879--0.923) & -- & 0.310 (0.266--0.358) & -- & 0.900 (0.881--0.919) & -- \\
DiffTumor \cite{chen2024towards} & 0.441 (0.374--0.493) & $+2.0$* & 0.899 (0.877--0.919) & $-0.3$ & 0.297 (0.251--0.342) & $-1.4$ & 0.899 (0.880--0.919) & $-0.1$ \\
DiffUltra SWAP & 0.451 (0.373--0.515) & $+3.0$* & 0.895 (0.871--0.917) & $-0.7$ & 0.300 (0.257--0.353) & $-1.1$ & 0.898 (0.877--0.916) & $-0.3$ \\
DiffUltra INSERT & \textbf{0.469 (0.376--0.542)} & $\mathbf{+4.8}$* & \textbf{0.904 (0.880--0.927)} & $\mathbf{+0.2}$ & \textbf{0.312 (0.267--0.359)} & $\mathbf{+0.2}$ & \textbf{0.901 (0.881--0.919)} & $0.0$ \\
\dotrule
\multicolumn{9}{@{}l}{\textbf{\textit{Average across scales:}}} \\
Real only & 0.340 (0.281--0.388) & -- & 0.853 (0.829--0.876) & -- & 0.233 (0.198--0.275) & -- & 0.871 (0.854--0.889) & -- \\
DiffTumor \cite{chen2024towards} & 0.362 (0.296--0.414) & $+2.2$* & 0.859 (0.837--0.882) & $+0.6$ & 0.226 (0.193--0.263) & $-0.7$ & 0.873 (0.855--0.892) & $+0.2$ \\
DiffUltra SWAP & 0.355 (0.282--0.415) & $+1.5$ & 0.845 (0.819--0.869) & $-0.8$ & 0.215 (0.184--0.250) & $-1.8$ & 0.865 (0.844--0.883) & $-0.7$ \\
DiffUltra INSERT & \textbf{0.385 (0.302--0.455)} & $\mathbf{+4.5}$* & \textbf{0.862 (0.837--0.886)} & $\mathbf{+0.9}$* & \textbf{0.241 (0.206--0.280)} & $\mathbf{+0.8}$ & \textbf{0.883 (0.864--0.901)} & $\mathbf{+1.1}$* \\
\hline\hline
\end{tabular*}
\vspace{-2pt}
\end{table*}

\section{EXPERIMENTAL SETTINGS}\label{experimental-settings}

We evaluate DiffUltra at multiple real-data scales to characterize how synthetic augmentation utility changes with training-set size. Experiments use 10\%, 50\%, and 100\% of the real training data, with augmentation volume defined relative to pathology-positive training frames at each scale.

\subsection{Datasets}\label{datasets}

The autoencoder is pretrained on all labeled and unlabeled data in the training partition, comprising 5.2M frames from 2,110 patients across 19 clinical sites. The diffusion model, skeleton dictionary, synthetic image filter (SIF), and downstream detectors use a fully annotated cohort of 598 patients across 12 sites, comprising 719,997 pleural effusion frames and 746,052 consolidation frames. Patients ranged from 0 to 90 years of age (mean $\pm$ SD: 47.3 $\pm$ 25.1); 42.3\% were female and 57.7\% male. Table~\ref{tab:2} summarizes dataset composition.

Video-level ground truth was established by majority vote among three independent LUS-trained emergency physicians. Trained researchers annotated frame-level bounding boxes for pleural effusion and consolidation, which were reviewed by at least one physician. Training, validation, and test partitions were split at the subject level. All data were collected under IRB-approved protocols with written informed consent.

\subsection{Implementation Details}\label{implementation-details}

\textbf{Autoencoder:} A VQGAN \cite{esser2021taming} encodes $512\times512$ frames into $64\times64\times8$ latent features \cite{chou2025ultrasound}. Training uses L1 reconstruction, structural similarity index measure (SSIM) \cite{wang2004ssim}, learned perceptual image patch similarity (LPIPS) \cite{zhang2018unreasonable}, KL loss, and patch-based adversarial loss \cite{isola2017image}. The checkpoint with the lowest total validation loss is selected.

\textbf{Conditional latent diffusion model:} Following \cite{chou2025ultrasound}, the denoising network uses a Stable Diffusion-style backbone \cite{rombach2022high} as in MedFusion \cite{mullerfranzes2023multimodal}. The denoiser is trained for 100 epochs with batch size 6 using the noise-prediction objective (Eq.~\eqref{eq:diffusion_loss}) with a scaled-linear noise schedule ($\beta_{\mathrm{start}}=0.002$, $\beta_{\mathrm{end}}=0.02$). The inference checkpoint is selected based on foreground structural similarity and image mean squared error on the validation set. DDIM sampling uses 150 steps.

\textbf{Structural representation and contextual positioning:} The region of interest $R$ is cropped from each annotated bounding box. For INSERT, the spatial binning function $\psi(\cdot)$ discretizes $(\Delta x,\Delta y,x,y)$ into an $N\times N\times N\times N$ grid ($N=10$) to estimate the PMF $P(\Delta X,\!\Delta Y|X,\!Y)$. The anatomical landmark center $(x,y)$ is obtained from annotated pleural-line boxes. For SWAP, the candidate set uses size tolerance $\rho=1.3$ and Dice threshold $\tau_{\text{Dice}}=0.2$. For the skeleton mask \(m\), echogenicity values are $v_1=0.24$, $v_2=0.5$, and $v_3=1$.

\textbf{Task-based SIF and downstream detection model:} Each detector is a separately trained off-the-shelf YOLOv5n model (2.17M parameters, 3.3 GFLOPs) \cite{jocher2020yolov5}, selected for its real-time detection capability \cite{terven2023comprehensive,ragab2024comprehensive} and lightweight architecture, which supports deployment on resource-constrained ultrasound hardware \cite{wu2021cachetrack,bottino2025automatic}. Detectors are trained for 300 epochs at $512\times512$ resolution using Adam (learning rate 0.001), a linear learning-rate schedule, and default augmentation settings and hyperparameters \cite{jocher2020yolov5}. The SIF detector is trained exclusively on real images; the downstream detector uses the same real images plus filtered synthetic images. All experiments use $\tau_{\text{conf}}=0.1$ and $\tau_{\text{IoU}}=0.01$. Models are implemented in PyTorch and trained on NVIDIA L40S GPUs.

\subsection{Evaluation Protocol and Metrics}\label{evaluation-protocol-and-metrics}

Frame-level detection is evaluated using average precision at IoU 0.5 (AP@0.5) \cite{lin2014microsoft}. Video-level classification uses AUROC, with the $k$-th highest frame-level detection confidence as the video score ($k=8$) \cite{li2025spatiotemporal}. Each augmentation strategy (INSERT, SWAP, and DiffTumor \cite{chen2024towards}) is compared with real-only training at the same scale; INSERT is also compared with DiffTumor. Differences ($\Delta$) are reported in percentage points. 95\% confidence intervals (CIs) and two-sided $p$-values are estimated by paired patient-level bootstrap (2,000 iterations) \cite{field2007bootstrapping}. Pooled effects are the unweighted mean of per-scale differences computed within each bootstrap iteration.

\subsection{Baseline Comparisons and Ablation Experiments}\label{baseline-comparisons-and-ablation-experiments}

DiffUltra is compared with real-only training and with DiffTumor \cite{chen2024towards}, chosen as the most closely related lesion-insertion method that has demonstrated gains in downstream tumor detection. SIF is applied only to DiffUltra, as DiffTumor has no task-based filtering step. For the ultrasound setting, DiffTumor is adapted by replacing boundary-based placement with the INSERT pathology position sampler and tumor masks with binary skeleton masks that provide basic structural guidance for ultrasound findings. Ablations at the 100\% real-data scale assess the contributions of the structural mask, conditional-PMF positional guidance, and task-based filtering. All synthetic-data experiments use $\gamma_{\text{syn}}=1$.

% ===================== Figure V =====================
\begin{figure*}[!t]
\centering
\includegraphics[width=\textwidth]{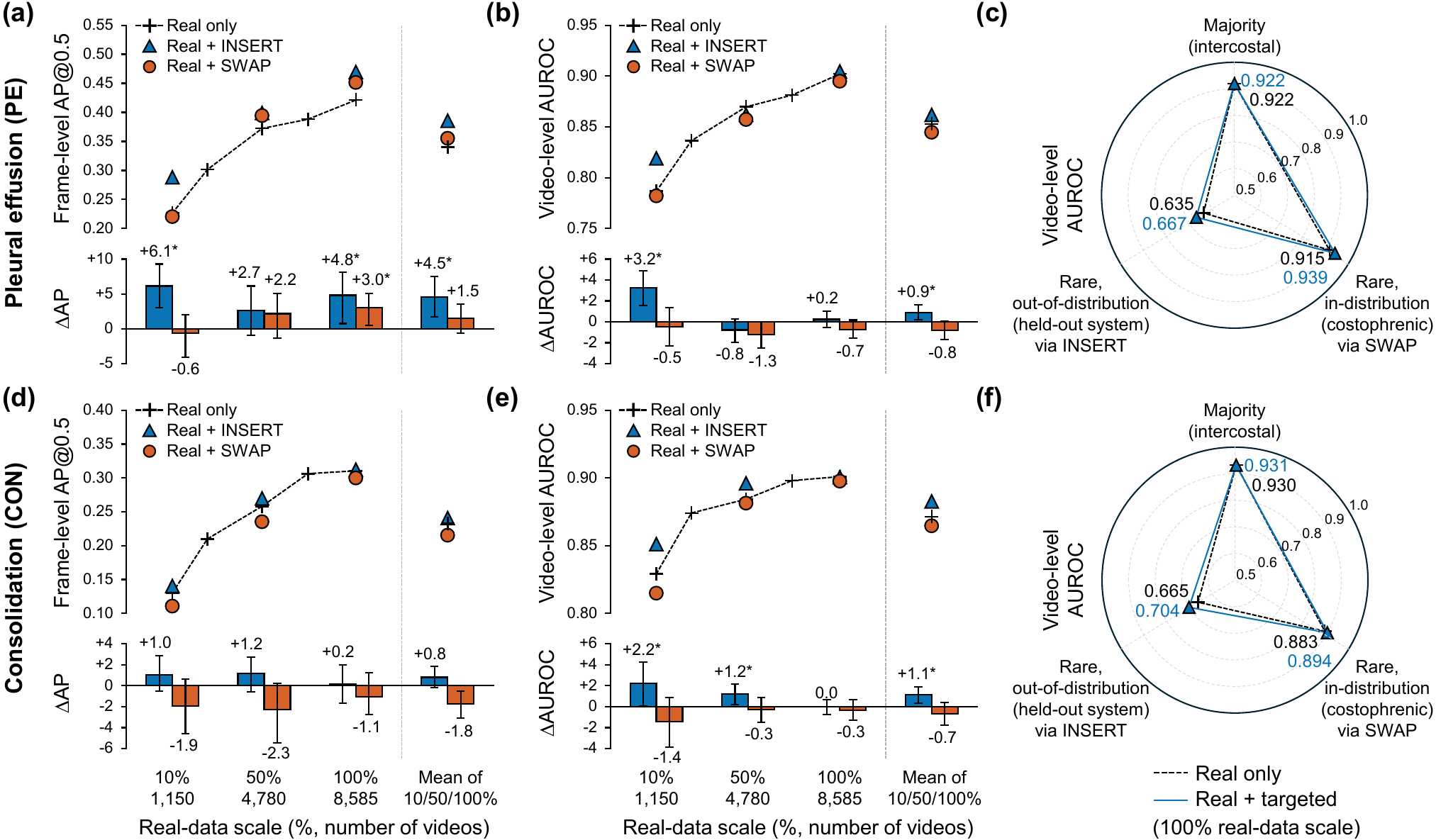}
\caption{Detection performance across real-data scales for pleural effusion (PE, top row) and consolidation (CON, bottom row). (a, d) Frame-level AP@0.5 and (b, e) video-level AUROC for real-only, INSERT-, and SWAP-augmented training; real-only models were also trained at 25\% and 75\%. Bars show changes from real-only in percentage points, with paired patient-level bootstrap 95\% CIs; *$p<0.05$, two-sided. ``Mean'' is the unweighted average across 10\%, 50\%, and 100\% scales. (c, f) Video-level AUROC at 100\% scale for the majority intercostal group and two rare subgroups, with and without targeted synthesis.}
\label{fig:main_results}
\end{figure*}

\subsection{Targeted Subgroup Experiments}\label{targeted-subgroup-experiments}

Targeted synthesis is evaluated for two data coverage gaps. First, costophrenic views account for 7.4\% of training data, compared with 92.6\% for intercostal views, and have a higher prevalence of severe pathology. Targeted SWAP augmentation is applied exclusively to costophrenic training data and evaluated on a held-out costophrenic subgroup. Second, an acquisition system absent from training and validation serves as an out-of-distribution subgroup. INSERT places structural representations from the training set onto pathology-absent backgrounds from this system, taken from patients outside the test subgroup. Both targeted experiments use $\gamma_{\text{syn}}=1$.

\section{Results}\label{results}

\subsection{Detection Performance Across Real-Data Scales}\label{detection-performance-across-real-data-scales}

Performance when training on real-only data was pathology-dependent (Fig.~\ref{fig:main_results} and Table~\ref{tab:synthetic_utility_combined_ioup5}). PE AP increased steadily with data scale, from 0.227 at 10\% to 0.372 at 50\% and 0.421 at 100\% scale. CON AP rose sharply from 0.130 at 10\% to 0.258 at 50\%, then more gradually to 0.310 at 100\%.

%?The effects of synthetic augmentation varied by pathology and real-data scale. For PE, INSERT improved AP over the real-only baseline at every scale, by 6.1, 2.7, and 4.8 points at 10\%, 50\%, and 100\%, respectively. The gains were significant at 10\% and 100\% ($p < 0.01$ and $p = 0.018$), with an average improvement of 4.5 points across scales (95\% CI 1.7--7.5; $p < 0.01$). PE video-level AUROC also increased at 10\% (+3.2 points; $p < 0.01$). For CON, INSERT improved AP at all three scales (1.0, 1.2, and 0.2 points), although not significantly. Video-level AUROC increased significantly at 10\% (+2.2 points; $p = 0.048$) and 50\% (+1.2 points; $p = 0.025$), with an average gain of 1.1 points across scales (95\% CI 0.3--1.9; $p < 0.01$). The benefit of SWAP depended on pathology. For PE, it improved AP at 50\% and 100\% (+2.2 and +3.0 points), with an average increase of 1.5 points. For CON, SWAP did not improve over the real-only baseline.
The effects of synthetic augmentation varied by pathology and real-data scale. Averaged across scales, INSERT significantly improved video-level AUROC over the real-only baseline for both PE (0.9 points; $p = 0.017$) and CON (1.1 points; $p < 0.01$). INSERT also improved AP by an average of 4.5 points for PE ($p < 0.01$) and 0.8 points for CON (n.s.).

For PE, the improvement in AP using INSERT was observed at every data scale, by 6.1, 2.7, and 4.8 points at 10\%, 50\%, and 100\%, respectively. INSERT also improved CON AP at all scales (1.0, 1.2, and 0.2 points), although this did not reach significance. The largest AUROC gains occurred at the small (10\%) scale (3.2 points for PE, $p < 0.01$; 2.2 points for CON, $p = 0.048$). The benefit of SWAP was more variable. For PE, SWAP improved AP at 50\% and 100\% (2.2 and 3.0 points), but not AUROC. For CON, SWAP numerically underperformed the real-only baseline, though differences were not significant.

Compared with DiffTumor \cite{chen2024towards}, INSERT yielded larger AP gains for both pathologies at every scale, despite retaining fewer images after SIF filtering (Table~\ref{tab:synthetic_utility_combined_ioup5}). For PE, both methods improved AP at every scale, but INSERT outperformed DiffTumor by 2.7, 1.4, and 2.8 points at 10\%, 50\%, and 100\% (2.3 points on average; $p = 0.065$). For CON, DiffTumor remained below the real-only baseline in AP at all scales, whereas INSERT consistently exceeded it. Averaged across scales, INSERT outperformed DiffTumor by 1.5 AP points ($p < 0.01$) and 0.9 AUROC points ($p = 0.011$).

\subsection{Ablation of Pipeline Components}\label{ablation-of-pipeline-components}

Table~\ref{tab:ablation_study_short} isolates the contributions of the three proposed components using a leave-one-component-out design: real data alone (row 1), synthetic data without DiffUltra refinements (row 2), configurations with one component removed (rows 3--5), and the full pipeline (row 6). All ablations use PE at full real-data scale with \(\gamma_{\text{syn}}=1\).

Adding synthetic data without structural, positional, or SIF guidance produced only a marginal gain (row 2: 0.2 AP points over baseline). The full pipeline achieved the highest AP (row 6: 4.8 points over baseline). Removing structural representation caused the largest reduction (4.4 points below the full pipeline; row 5), followed by removing positional guidance (2.7 points below the full pipeline; row 3) and SIF filtering (2.5 points below the full pipeline; row 4). These results suggest that all three components contribute to performance, with the full pipeline achieving more than twice the gain of any configuration with one component removed.

\subsection{Targeted Subgroup Performance}\label{targeted-subgroup-performance}

Table~\ref{tab:id_ood_performance} reports video-level AUROC at full real-data scale for two coverage gaps. Each real-only model is compared with its counterpart trained using targeted synthetic augmentation.

For the in-distribution costophrenic zone, which accounted for 7.4\% of training data, targeted SWAP increased AUROC by 2.4 points for PE and 1.1 for CON. In the non-targeted intercostal majority, AUROC changes were marginal (0.0 for PE, $-0.1$ for CON), indicating that gains in the minority subgroup were achieved without affecting majority performance. For the out-of-distribution held-out acquisition system, targeted INSERT increased AUROC by 3.2 for PE and 3.9 for CON, even though the system was absent from both training and validation. Together, these findings suggest that targeted synthesis can improve performance at full real-data scale for both an underrepresented in-distribution anatomical subgroup and an out-of-distribution acquisition system.

\section{Discussion}\label{discussion}

% ===================== Table III =====================
\begin{table}[!t]
\centering
\footnotesize
\caption{Ablations for PE at 100\% real-data scale with $\gamma_{\text{syn}}=1$. $m$: tri-level mask; Position: positional guidance $P(\Delta X,\!\Delta Y|X,\!Y)$; SIF: synthetic image filter. \checkmark: used; X: not used; --: no synthetic data. AP@0.5 with patient-level bootstrap 95\% CIs; $\Delta$ is the change in AP@0.5 from the row 1 baseline.}
\label{tab:ablation_study_short}
\setlength{\tabcolsep}{4pt}
\renewcommand{\arraystretch}{1.08}
\begin{tabular*}{\columnwidth}{@{\extracolsep{\fill}}cccccc@{}}
\hline\hline
\textbf{Exp.} & $\boldsymbol{m}$ & \textbf{Position} & \textbf{SIF} & \textbf{AP@0.5 (95\% CI)} & \textbf{$\Delta$} \\
\midrule
1 & -- & -- & -- & 0.421 (0.358--0.473) & baseline \\
\dotrule
2 & X & X & X & 0.423 (0.359--0.475) & $+0.2$ \\
3 & $\checkmark$ & X & $\checkmark$ & 0.442 (0.363--0.505) & $+2.1$ \\
4 & $\checkmark$ & $\checkmark$ & X & 0.444 (0.365--0.500) & $+2.3$ \\
5 & X & $\checkmark$ & $\checkmark$ & 0.425 (0.361--0.481) & $+0.4$ \\
6 & $\checkmark$ & $\checkmark$ & $\checkmark$ & \textbf{0.469 (0.376--0.542)} & $\mathbf{+4.8}$ \\
\hline\hline
\end{tabular*}
\end{table}

% ===================== Table IV =====================
\begin{table}[!t]
\centering
\footnotesize
\caption{Video-level AUROC with targeted synthetic augmentation at 100\% real-data scale. Intercostal videos form the in-distribution majority and costophrenic videos form the minority; the held-out acquisition system was absent from training and validation. Pos.: pathology-positive videos (\% of subgroup). $\Delta$: change from real-only, in percentage points.}
\label{tab:id_ood_performance}
\setlength{\tabcolsep}{2pt}
\renewcommand{\arraystretch}{1.08}
\setlength{\aboverulesep}{1pt}\setlength{\belowrulesep}{1pt}
\begin{tabular*}{\columnwidth}{@{\extracolsep{\fill}}llrrccc@{}}
\hline\hline
\textbf{Finding}
& \textbf{Subgroup}
& \textbf{Videos}
& \textbf{Pos. (\%)}
& \multicolumn{2}{c}{\textbf{AUROC}}
& \textbf{$\Delta$} \\[-1pt]
\cmidrule(lr){5-6}
& & & & \textbf{Real only} & \textbf{Real+Syn.} & \\[-1pt]
\midrule
\multicolumn{7}{@{}l}{\textbf{In-distribution: SWAP augmentation}} \\[-1pt]
PE  & Intercostal  & 3601 & 470 (13.1) & 0.922 & 0.922 & $0.0$ \\
PE  & Costophrenic & 332  & 74 (22.3)  & 0.915 & \textbf{0.939} & \textbf{+2.4} \\
CON & Intercostal  & 3611 & 561 (15.5) & 0.931 & 0.930 & $-0.1$ \\
CON & Costophrenic & 343  & 66 (19.2)  & 0.883 & \textbf{0.894} & \textbf{+1.1} \\
\dotrule
\multicolumn{7}{@{}l}{\textbf{Out-of-distribution: INSERT augmentation}} \\[-1pt]
PE  & Held-out system & 212 & 21 (9.9) & 0.635 & \textbf{0.667} & \textbf{+3.2} \\
CON & Held-out system & 213 & 13 (6.1) & 0.665 & \textbf{0.704} & \textbf{+3.9} \\
\hline\hline
\end{tabular*}
\vspace{-5pt}
\end{table}

DiffUltra addresses foreground--background compatibility in ultrasound pathology synthesis in three stages: structural representation via a tri-level conditioning mask, contextual positioning via INSERT and SWAP retrieval, and synthetic image filtering to assess downstream utility. Synthetic insertion improved detection of both pathologies on average across real-data scales, with the largest gains when real training data were scarce. The benefit was pathology-dependent, with larger PE gains that persisted at full scale and smaller CON gains concentrated at low and intermediate scales. This contrast suggests that synthetic-data utility should not be assumed uniform across pathologies. One possible explanation is greater heterogeneity in PE presentation, e.g., variation in fluid volume and loculation, although this was not directly tested. With DiffTumor, binary-mask insertion improved AP for PE but not CON, whereas INSERT improved AP for both pathologies. This pattern is consistent with the tri-level structural representation being particularly valuable for consolidation, whose mixed hypoechoic and hyperechoic content a binary mask cannot capture. These findings support pathology-specific evaluation of synthetic augmentation.

INSERT and SWAP also behaved differently. INSERT improved AP at every scale for both pathologies, whereas SWAP improved AP for PE but not for CON (Table~\ref{tab:synthetic_utility_combined_ioup5}). Because SWAP replaces an existing finding, it primarily varies pathology appearance rather than anatomical context, and pathologies may differ in their sensitivity to this distinction. SWAP nevertheless improved the targeted costophrenic subgroup (Table~\ref{tab:id_ood_performance}), suggesting that the two strategies address different coverage gaps.

Because each synthetic image is screened against the intended detections, the same framework supports both general and targeted synthesis. The costophrenic and held-out-system experiments apply identical structural representation, positioning guidance, and task-based filtering criteria to purposefully selected subsets. Targeted synthesis produced additional performance gains for these subgroups at full data scale, indicating that coverage gaps can remain even when aggregate training data are abundant. These experiments also illustrate the value of assessing performance across clinically relevant anatomical and acquisition subgroups rather than relying only on aggregate metrics \cite{fda2025ai}. Targeted synthesis may help address identified coverage gaps when additional positive cases from the target subgroup are difficult to obtain.

Several limitations should be considered. First, some targeted subgroups are modest in size (13--74 positive cases), so these effect estimates are less precise than those from the full evaluation cohort and require confirmation in larger, independent cohorts before clinical use. Second, SIF shares its architecture with the downstream detector and is trained for a similar task, so the models may share failure modes, allowing systematic errors in synthetic images to pass filtering. This risk may be greater in targeted subgroups, where a small number of undetected failures could disproportionately affect performance. Filtering with ensembles of detectors, and accepting only images on which they agree, may reduce this risk \cite{dietterich2000ensemble}, \cite{solovyev2021weighted}, as may expert review of accepted images.

Third, all experiments used a fixed synthetic-to-real ratio (\(\gamma_{\text{syn}}=1\)), a single detector architecture (YOLOv5n) \cite{jocher2020yolov5} for SIF and downstream training, and a single autoencoder and diffusion backbone. YOLOv5n was chosen because it is a widely used real-time detector \cite{terven2023comprehensive}, \cite{ragab2024comprehensive} suitable for deployment on resource-constrained ultrasound hardware \cite{wu2021cachetrack}, \cite{bottino2025automatic}. The magnitude of the gains may depend on these choices, and the optimal synthesis ratio may vary by pathology and data scale. Fourth, comparison was limited to DiffTumor \cite{chen2024towards}, the most closely related mask-conditioned lesion-insertion method. GAN-based ultrasound synthesis \cite{liang2022sketch}, \cite{fatima2025synthetic} and copy-paste augmentation \cite{ghiasi2021simple} were not evaluated. Copy-paste transfers a finding's original pixels without adapting them to the host anatomy, whereas DiffUltra explicitly addresses this compatibility problem; for GAN-based synthesis, latent diffusion models have compared favorably with GANs in medical image synthesis \cite{mullerfranzes2023multimodal}. Both are useful reference points for future comparative studies. Fifth, the experiments focus on CON and PE in LUS; generalization to other pathologies, anatomical regions, imaging modalities, acquisition systems, and anatomical references beyond the pleural line (e.g., rib shadows) remains to be established.

\section{Conclusion}\label{conclusion}

We present DiffUltra, a latent diffusion framework for lung ultrasound pathology synthesis that addresses foreground--background compatibility through three stages: structural representation, contextual positioning, and task-based filtering. Synthetic insertion improved detection of both pathologies on average across training-set sizes, with the largest gains when real data were scarce. Targeted synthesis further improved performance in underrepresented anatomical and acquisition subgroups at full data scale. These findings suggest that task-guided synthesis is a promising strategy for addressing residual coverage gaps in ultrasound AI training data.

\section*{Acknowledgments}\label{acknowledgments}

The authors thank {C. L. Moore, C. Baloescu, R. Denkewicz, M. Hesami, B. Hicks, N. Schnittke, M. Parker, A. Silver, M. Petrovich, A. Sabbaj, L. Hu, D. Chan, M. Zhu, M. Kaili, C. Thomas, M. Feuerherdt, Y. Zhang, C. Gregory, K. Gregory, D. O. Kessler, D. Coneybeare, L. Malia, J. Shupp, C. Mehanian, S. Kulhare, and R. Millin} for data acquisition and annotation.

\bibliographystyle{IEEEtran}
\bibliography{references}

\begin{thebibliography}{10}
\providecommand{\url}[1]{#1}
\csname url@samestyle\endcsname
\providecommand{\newblock}{\relax}
\providecommand{\bibinfo}[2]{#2}
\providecommand{\BIBentrySTDinterwordspacing}{\spaceskip=0pt\relax}
\providecommand{\BIBentryALTinterwordstretchfactor}{4}
\providecommand{\BIBentryALTinterwordspacing}{\spaceskip=\fontdimen2\font plus
\BIBentryALTinterwordstretchfactor\fontdimen3\font minus \fontdimen4\font\relax}
\providecommand{\BIBforeignlanguage}[2]{{%
\expandafter\ifx\csname l@#1\endcsname\relax
\typeout{** WARNING: IEEEtran.bst: No hyphenation pattern has been}%
\typeout{** loaded for the language `#1'. Using the pattern for}%
\typeout{** the default language instead.}%
\else
\language=\csname l@#1\endcsname
\fi
#2}}
\providecommand{\BIBdecl}{\relax}
\BIBdecl

\bibitem{oakdenrayner2020hidden}
L.~Oakden-Rayner, J.~Dunnmon, G.~Carneiro, and C.~R{\'e}, ``Hidden stratification causes clinically meaningful failures in machine learning for medical imaging,'' in \emph{Proc. ACM Conf. Health, Inference, Learn. ({CHIL})}, 2020, pp. 151--159.

\bibitem{salmi2024handling}
M.~Salmi, D.~Atif, D.~Oliva, A.~Abraham, and S.~Ventura, ``Handling imbalanced medical datasets: Review of a decade of research,'' \emph{Artif. Intell. Rev.}, vol.~57, no.~10, 2024, art. no. 273.

\bibitem{chawla2002smote}
N.~V. Chawla, K.~W. Bowyer, L.~O. Hall, and W.~P. Kegelmeyer, ``{SMOTE}: Synthetic minority over-sampling technique,'' \emph{J. Artif. Intell. Res.}, vol.~16, pp. 321--357, Jun. 2002.

\bibitem{goodfellow2014gan}
I.~Goodfellow \emph{et~al.}, ``Generative adversarial nets,'' in \emph{Proc. Adv. Neural Inf. Process. Syst.}, vol.~27, 2014, pp. 2672--2680.

\bibitem{ho2020denoising}
J.~Ho, A.~Jain, and P.~Abbeel, ``Denoising diffusion probabilistic models,'' in \emph{Proc. Adv. Neural Inf. Process. Syst.}, vol.~33, 2020, pp. 6840--6851.

\bibitem{ktena2024generative}
I.~Ktena \emph{et~al.}, ``Generative models improve fairness of medical classifiers under distribution shifts,'' \emph{Nature Med.}, vol.~30, no.~4, pp. 1166--1173, Apr. 2024.

\bibitem{xing2023you}
X.~Xing, F.~Felder, Y.~Nan, G.~Papanastasiou, S.~Walsh, and G.~Yang, ``You don't have to be perfect to be amazing: Unveil the utility of synthetic images,'' in \emph{Proc. Med. Image Comput. Comput.-Assist. Intervent. ({MICCAI})}, 2023, pp. 13--22.

\bibitem{chen2024towards}
Q.~Chen \emph{et~al.}, ``Towards generalizable tumor synthesis,'' in \emph{Proc. IEEE/CVF Conf. Comput. Vis. Pattern Recognit. ({CVPR})}, 2024, pp. 11\,147--11\,158.

\bibitem{yang2022current}
T.~Yang, O.~Karaku{\c{s}}, N.~Anantrasirichai, and A.~Achim, ``Current advances in computational lung ultrasound imaging: A review,'' \emph{IEEE Trans. Ultrason., Ferroelectr., Freq. Control}, vol.~70, no.~1, pp. 2--15, Jan. 2023.

\bibitem{chou2025ultrasound}
Y.-C. Chou \emph{et~al.}, ``Ultrasound image synthesis using generative {AI} for lung consolidation detection,'' in \emph{Proc. IEEE 22nd Int. Symp. Biomed. Imag. ({ISBI})}, 2025, pp. 1--5.

\bibitem{roy2020deep}
S.~Roy \emph{et~al.}, ``Deep learning for classification and localization of {COVID-19} markers in point-of-care lung ultrasound,'' \emph{IEEE Trans. Med. Imag.}, vol.~39, no.~8, pp. 2676--2687, Aug. 2020.

\bibitem{tsai2021automatic}
C.-H. Tsai \emph{et~al.}, ``Automatic deep learning-based pleural effusion classification in lung ultrasound images for respiratory pathology diagnosis,'' \emph{Phys. Med.}, vol.~83, pp. 38--45, Mar. 2021.

\bibitem{durrani2022automatic}
N.~Durrani \emph{et~al.}, ``Automatic deep learning-based consolidation/collapse classification in lung ultrasound images for {COVID-19} induced pneumonia,'' \emph{Sci. Rep.}, vol.~12, 2022, art. no. 17581.

\bibitem{li2025spatiotemporal}
G.~Y. Li \emph{et~al.}, ``Spatiotemporal learning with context-aware video tubelets for ultrasound video analysis,'' in \emph{Proc. IEEE 22nd Int. Symp. Biomed. Imag. ({ISBI})}, 2025, pp. 1--5.

\bibitem{zhang2025lefusion}
H.~Zhang \emph{et~al.}, ``{LeFusion}: Controllable pathology synthesis via lesion-focused diffusion models,'' in \emph{Proc. Int. Conf. Learn. Represent. ({ICLR})}, 2025.

\bibitem{liang2022sketch}
J.~Liang \emph{et~al.}, ``Sketch guided and progressive growing {GAN} for realistic and editable ultrasound image synthesis,'' \emph{Med. Image Anal.}, vol.~79, 2022, art. no. 102461.

\bibitem{stojanovski2023echo}
D.~Stojanovski, U.~Hermida, P.~Lamata, A.~Beqiri, and A.~Gomez, ``Echo from noise: Synthetic ultrasound image generation using diffusion models for real image segmentation,'' in \emph{Proc. Int. Workshop Adv. Simplifying Med. Ultrasound ({ASMUS})}, 2023, pp. 34--43.

\bibitem{dahan2025csg}
E.~Dahan \emph{et~al.}, ``{CSG}: A context-semantic guided diffusion approach in de novo musculoskeletal ultrasound image generation,'' in \emph{Proc. 47th Annu. Int. Conf. IEEE Eng. Med. Biol. Soc. ({EMBC})}, 2025, pp. 1--7.

\bibitem{duan2025fetalflex}
Y.~Duan \emph{et~al.}, ``{FetalFlex}: Anatomy-guided diffusion model for flexible control on fetal ultrasound image synthesis,'' \emph{Med. Image Anal.}, vol. 105, Oct. 2025, art. no. 103725.

\bibitem{heidari2026awdiff}
M.~Heidari, N.~Anantrasirichai, S.~Walker, R.~Bhatnagar, and A.~Achim, ``{AWDiff}: An a trous wavelet diffusion model for lung ultrasound image synthesis,'' in \emph{Proc. IEEE Int. Conf. Acoust., Speech, Signal Process. ({ICASSP})}, 2026, pp. 6891--6895.

\bibitem{zhang2023diffusion}
X.~Zhang, A.~Gangopadhyay, H.-M. Chang, and R.~Soni, ``Diffusion model-based data augmentation for lung ultrasound classification with limited data,'' in \emph{Proc. Mach. Learn. Health ({ML4H})}, ser. Proc. Mach. Learn. Res., vol. 225, 2023, pp. 664--676.

\bibitem{stojanovski2024efficient}
D.~Stojanovski, M.~da~Silva, P.~Lamata, A.~Beqiri, and A.~Gomez, ``Efficient semantic diffusion architectures for model training on synthetic echocardiograms,'' arXiv:2409.19371, 2024.

\bibitem{fatima2025synthetic}
N.~Fatima \emph{et~al.}, ``Synthetic lung ultrasound data generation using autoencoder with generative adversarial network,'' \emph{IEEE Trans. Ultrason., Ferroelectr., Freq. Control}, vol.~72, no.~5, pp. 624--635, May 2025.

\bibitem{tian2025enhancing}
Y.~Tian, E.~Ucurum, X.~Han, R.~Young, C.~Chatwin, and P.~Birch, ``Enhancing fetal plane classification accuracy with data augmentation using diffusion models,'' \emph{IET Image Process.}, vol.~19, no.~1, 2025, art. no. e70151.

\bibitem{nazir2025diffusion}
M.~Nazir, M.~Aqeel, and F.~Setti, ``Diffusion-based data augmentation for medical image segmentation,'' in \emph{Proc. IEEE/CVF Int. Conf. Comput. Vis. ({ICCV}) Workshops}, 2025, pp. 1341--1350.

\bibitem{gatoula2026interpretable}
P.~Gatoula, G.~Dimas, and D.~K. Iakovidis, ``Interpretable similarity of synthetic image utility,'' \emph{IEEE Trans. Med. Imag.}, vol.~45, no.~7, pp. 3529--3544, Jul. 2026.

\bibitem{frid2018gan}
M.~Frid-Adar, I.~Diamant, E.~Klang, M.~Amitai, J.~Goldberger, and H.~Greenspan, ``{GAN}-based synthetic medical image augmentation for increased {CNN} performance in liver lesion classification,'' \emph{Neurocomputing}, vol. 321, pp. 321--331, Dec. 2018.

\bibitem{ghiasi2021simple}
G.~Ghiasi \emph{et~al.}, ``Simple copy-paste is a strong data augmentation method for instance segmentation,'' in \emph{Proc. IEEE/CVF Conf. Comput. Vis. Pattern Recognit. ({CVPR})}, 2021, pp. 2917--2927.

\bibitem{chen2025scaling}
Q.~Chen \emph{et~al.}, ``Scaling tumor segmentation: Best lessons from real and synthetic data,'' in \emph{Proc. IEEE/CVF Int. Conf. Comput. Vis. ({ICCV})}, 2025, pp. 24\,001--24\,013.

\bibitem{rombach2022high}
R.~Rombach, A.~Blattmann, D.~Lorenz, P.~Esser, and B.~Ommer, ``High-resolution image synthesis with latent diffusion models,'' in \emph{Proc. IEEE/CVF Conf. Comput. Vis. Pattern Recognit. ({CVPR})}, 2022, pp. 10\,684--10\,695.

\bibitem{song2021ddim}
J.~Song, C.~Meng, and S.~Ermon, ``Denoising diffusion implicit models,'' in \emph{Proc. Int. Conf. Learn. Represent. ({ICLR})}, 2021.

\bibitem{otsu1975threshold}
N.~Otsu, ``A threshold selection method from gray-level histograms,'' \emph{IEEE Trans. Syst., Man, Cybern.}, vol. SMC-9, no.~1, pp. 62--66, Jan. 1979.

\bibitem{esser2021taming}
P.~Esser, R.~Rombach, and B.~Ommer, ``Taming transformers for high-resolution image synthesis,'' in \emph{Proc. IEEE/CVF Conf. Comput. Vis. Pattern Recognit. ({CVPR})}, 2021, pp. 12\,873--12\,883.

\bibitem{wang2004ssim}
Z.~Wang, A.~C. Bovik, H.~R. Sheikh, and E.~P. Simoncelli, ``Image quality assessment: From error visibility to structural similarity,'' \emph{IEEE Trans. Image Process.}, vol.~13, no.~4, pp. 600--612, Apr. 2004.

\bibitem{zhang2018unreasonable}
R.~Zhang, P.~Isola, A.~A. Efros, E.~Shechtman, and O.~Wang, ``The unreasonable effectiveness of deep features as a perceptual metric,'' in \emph{Proc. IEEE/CVF Conf. Comput. Vis. Pattern Recognit. ({CVPR})}, 2018, pp. 586--595.

\bibitem{isola2017image}
P.~Isola, J.-Y. Zhu, T.~Zhou, and A.~A. Efros, ``Image-to-image translation with conditional adversarial networks,'' in \emph{Proc. IEEE Conf. Comput. Vis. Pattern Recognit. ({CVPR})}, 2017, pp. 1125--1134.

\bibitem{mullerfranzes2023multimodal}
G.~M{\"u}ller-Franzes \emph{et~al.}, ``A multimodal comparison of latent denoising diffusion probabilistic models and generative adversarial networks for medical image synthesis,'' \emph{Sci. Rep.}, vol.~13, 2023, art. no. 12098.

\bibitem{jocher2020yolov5}
\BIBentryALTinterwordspacing
G.~Jocher, ``Ultralytics {YOLOv5},'' Version 7.0, 2020. [Online]. Available: \url{https://github.com/ultralytics/yolov5}
\BIBentrySTDinterwordspacing

\bibitem{terven2023comprehensive}
J.~Terven, D.-M. C{\'o}rdova-Esparza, and J.-A. Romero-Gonz{\'a}lez, ``A comprehensive review of {YOLO} architectures in computer vision: From {YOLOv1} to {YOLOv8} and {YOLO-NAS},'' \emph{Mach. Learn. Knowl. Extr.}, vol.~5, no.~4, pp. 1680--1716, 2023.

\bibitem{ragab2024comprehensive}
M.~G. Ragab \emph{et~al.}, ``A comprehensive systematic review of {YOLO} for medical object detection (2018 to 2023),'' \emph{IEEE Access}, vol.~12, pp. 57\,815--57\,836, 2024.

\bibitem{wu2021cachetrack}
X.~Wu \emph{et~al.}, ``{CacheTrack-YOLO}: Real-time detection and tracking for thyroid nodules and surrounding tissues in ultrasound videos,'' \emph{IEEE J. Biomed. Health Inform.}, vol.~25, no.~10, pp. 3812--3823, 2021.

\bibitem{bottino2025automatic}
A.~Bottino \emph{et~al.}, ``Automatic approach for {B}-lines detection in lung ultrasound images using {You Only Look Once} algorithm,'' \emph{J. Ultrasound}, vol.~28, no.~4, pp. 985--992, 2025.

\bibitem{lin2014microsoft}
T.-Y. Lin \emph{et~al.}, ``Microsoft {COCO}: Common objects in context,'' in \emph{Proc. Eur. Conf. Comput. Vis. (ECCV)}, 2014, pp. 740--755.

\bibitem{field2007bootstrapping}
C.~A. Field and A.~H. Welsh, ``Bootstrapping clustered data,'' \emph{J. R. Stat. Soc. B, Stat. Methodol.}, vol.~69, no.~3, pp. 369--390, Jun. 2007.

\bibitem{fda2025ai}
\BIBentryALTinterwordspacing
{U.S. Food and Drug Administration}, ``Artificial intelligence-enabled device software functions: Lifecycle management and marketing submission recommendations,'' Draft Guidance, Jan. 2025. [Online]. Available: \url{https://www.fda.gov/media/184856/download}
\BIBentrySTDinterwordspacing

\bibitem{dietterich2000ensemble}
T.~G. Dietterich, ``Ensemble methods in machine learning,'' in \emph{Proc. Int. Workshop Multiple Classifier Syst. ({MCS})}, ser. Lecture Notes in Computer Science, vol. 1857, 2000, pp. 1--15.

\bibitem{solovyev2021weighted}
R.~Solovyev, W.~Wang, and T.~Gabruseva, ``Weighted boxes fusion: Ensembling boxes from different object detection models,'' \emph{Image Vis. Comput.}, vol. 107, 2021, art. no. 104117.

\end{thebibliography}

\end{document}